\documentclass [12pt]{article}
\usepackage {amssymb}
\usepackage {amsmath}
\usepackage {graphicx}

\begin{document}

\begin{flushleft}
\footnotesize{V. Krasnoholovets, S. Skliarenko and O. Strokach,
On the behavior of physical parameters of aqueous solutions affected by
the inerton field of Teslar$^{\footnotesize{\circledR}}$ technology,
{\it International Journal of Modern Physics B} {\bf 20}, no. 1, pp. 111-124 (2006) }
\end{flushleft}

\bigskip

\begin{center}
\subsection*{ON THE BEHAVIOR OF PHYSICAL PARAMETERS OF AQUEOUS SOLUTIONS
AFFECTED BY THE INERTON FIELD OF
TESLAR$^{\footnotesize{\circledR}}$  TECHNOLOGY}
\end{center}

\bigskip

\begin{center}
{\bf Volodymyr Krasnoholovets$^1$,   \\

\medskip

Sergiy Skliarenko$^2$ and  Olexander Strokach$^2$}
\end{center}

\bigskip

\begin{center}
{\small $^1$Department of Physics, Institute for Basic Research, \\
 90 East Winds Court, Palm Harbor, FL 34683, U.S.A.\\
e-mail address:  v\_kras@yahoo.com  \\}
\end{center}
\begin{center}
{\small $^2$Department of Receivers of Radiation, Institute of Physics, \\
National Academy of Sciences of Ukraine,\\
Prospekt Nauky 46, UA-03028 Kyiv, Ukraine \\
e-mail address: strokach@iop.kiev.ua}
\end{center}

\begin{abstract}
We present studies of the behavior of the permittivity of
such liquid systems as pure distilled water, alcohol and 50\%-aqueous
solutions of alcohol as affected by the inerton field generated by
a special signal generator contained within a wrist-watch or
bracelet made by so-called Teslar$^{\tiny{\circledR}}$ technology.
It has been found that the changes in fact are significant.
The method employed has allowed us to fix the value of frequency
of the field generated by the Teslar$^{\tiny{\circledR}}$ chip.
The frequency has been determined to be approximately 8 Hz.
The phenomenological consideration and submicroscopic
foundations of a significant increase of the permittivity
are studied taking into account an additional interaction,
namely the mass interaction between polar water molecules,
which is caused by the inerton field of the
Teslar$^{\tiny{\circledR}}$ chip. This is one more proof of
Krasnoholovets' concept regarding the existence of
a substructure of the matter waves of
moving/vibrating entities, i.e. the inerton field,
which has been predicted in a series of his previous works.

\medskip
\noindent
\textbf{key words:} {Teslar technology, permittivity, matter waves,
mass, inertons, \\ quantum mechanics}
\end{abstract}

\begin{flushright}
14 September 2005 \quad\quad \ \ \ \
\end{flushright}

\section{Introduction}

The influence of physical fields generated by sources of electromagnetic
waves of the so-called non-Hertzian type (scalar waves), has been marked in
operations of the medical and biologic profile [1-3] .
In those experiments, the effect of generators of scalar waves on
biological objects of various levels of organization was researched.
The \textit{Teslar}$^{\scriptsize{\circledR}}$
\textit{technology} is said to be of such a generator. Those authors
put forward the supposition that the effect of energy of scalar fields on
such nonlinear systems as biological objects was more essential than the
influence of conventional vector electromagnetic fields.

Those medical and biological experiments have allowed
other kinds of studies, namely, the examination of behavior
of chemical and physical systems affected by
the Teslar$^{\scriptsize{\circledR}}$ technology.
First of all such objects are liquids and crystals.
Taking into account the nonlinear behavior of responses
of biological objects to the Teslar$^{\scriptsize{\circledR}}$
chip, we have decided to examine those temperature regions
in which nonlinear properties of selected
objects are most clearly observed. It seems that
the first kind of phase transition of the system
studied (for instance, the liquid-steam transition)
is the most suitable for our purpose. That is why
in our experiments we have decided to examine features
of the influence of scalar fields of the
Teslar$^{\scriptsize{\circledR}}$ chip (the TC below)
on the process of evaporation of components of
the aqueous solution under examination.
More exactly, our purpose has been the comparison
of dielectric characteristics of aqueous solutions
of organic substances, both under and independent
of the influence of the TC. As model substances we
have taken distilled water, pure ethyl alcohol
C$_2$H$_5$OH, glycerin, and the aqueous solutions
of alcohol and glycerin. The most significant
results have been obtained during experiments
involving the 50 \%-aqueous solution of alcohol.

\section{EXPERIMENTAL}

\subsection{Experimental conditions}

Our experiments have been conducted in a special room shielded from
electromagnetic interference, in accordance with the National Standards of
Ukraine in support of unity of measurements. Namely, at such conditions,
measuring equipment yields results with accuracy up to 10 nV. This level of
measurement is quite sufficient for obtaining trustworthy information from
our experiments. The National Standards of Ukraine on support of unity of
measurements corresponds to the defined norms of international standard IEC
(International Electrotechnical Committee). Moreover, a grounded metal box
covered the cuvette with samples studied; the box was cube-shaped with equal
sides of approximately 12 cm.

In the room, the following common conditions were maintained:

- Barometric pressure was controlled between 750 to 770 mm of mercury
column,

- Temperature was maintained between 18 to 22 $^{\circ}$C,

- Relative humidity of air was maintained between 65 to 75 \%.

The experiments were conducted during normal day working hours.

\subsection{Measurements}

Two kinds of experiments have been performed:

\noindent
\textit{(a)} the study of the behavior of capacity of the 50\%-aqueous
solution of alcohol affected by the TC in the course of evaporation of the
solution components;

\noindent
\textit{(b)} the study of the behavior of capacity of the 50\%-aqueous
solution of alcohol affected by the TC and the modulated laser radiation in
the course of evaporation of the solution components.

\subsubsection{Measurements of scheme (\textit{ a})}

In the experiments of the kind \textit{(a)}
we used the set-up shown in Figure 1.

\begin{figure}
\begin{center}
\includegraphics[scale=0.65]{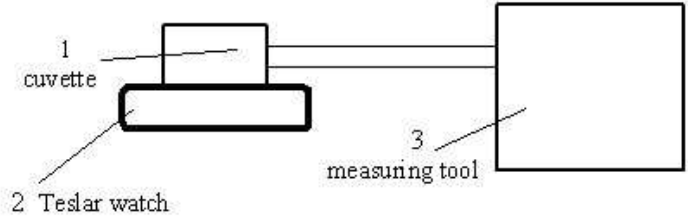}
\caption{\small{Scheme of the experiments of the kind \textit{(a)}.}} \label{Figure 1}
\end{center}
\end{figure}

The experiments were carried out with the use of a measuring cell,
i.e. the cuvette ``1'' with sizes $30 \times 4 \times 0.5$ mm$^{3}$.
It was a typical capacitor: two plates made of high-quality nickel,
which are jointed by thin teflon gaskets. The top surface of
the capacitor was open for free
evaporation of components of the solution. The capacity
of the aqueous solution was measured by device ``3''
that is the measuring tool of impedance E7-15. The value
of measuring field was equal to $U_{\rm meas}$ = 2 V;
the frequency of measuring field
was chosen equal to $f_{\rm meas}$ = 100 Hz
(for the first series of experiments) and $f_{\rm meas}$ = 1 kHz
(for the second series of experiments).

In the experiments we have investigated how the capacity of the solution
varies with time. We considered two cases: the aqueous solution without the
TC (control) and the aqueous solution affected by the TC (test samples). The
watch ``2'' has been placed as shown in Figure 1. The distance between the
watch and the cuvette was equal to 1mm.

The residual solution was weighed and its volume measured to estimate the
density.

The experimental results are presented in Figures 2 to 5.

\begin{figure}
\begin{center}
\includegraphics[scale=0.6]{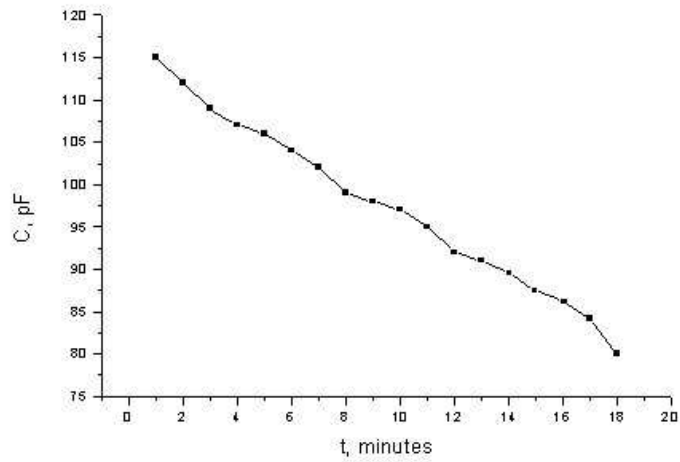}
\caption{\small{Capacity of the 50\%-aqueous solution of alcohol as a
function of time at the frequency of measuring electric field $f_{\rm meas}$ = 1
kHz without the influence of the TC. However, a conventional quartz watch
(an imitator) is used, under the cuvette, for the compensation of influence
of the metal case of the Teslar Watch on the allocation of the strength of
the measuring field in the experimental cell, i.e. cuvette.}} \label{Figure 2}
\end{center}
\end{figure}

\begin{figure}
\begin{center}
\includegraphics[scale=0.6]{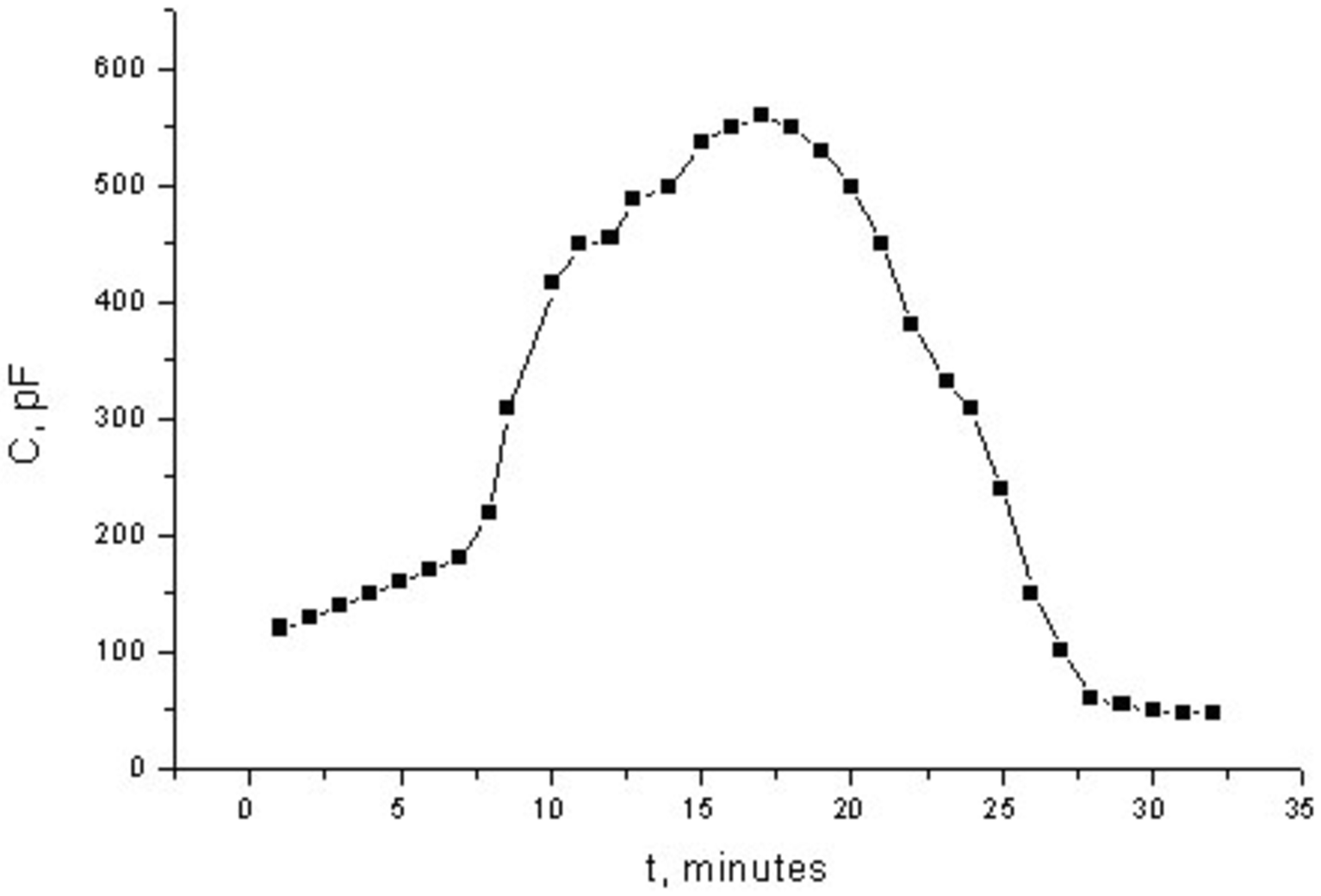}
\caption{\small{Capacity of the 50\%-aqueous solution of alcohol affected
by the TC as a function of time at the frequency of measuring electric field
$f_{\rm meas}$ = 1 kHz.}} \label{Figure 3}
\end{center}
\end{figure}

\begin{figure}
\begin{center}
\includegraphics[scale=0.6]{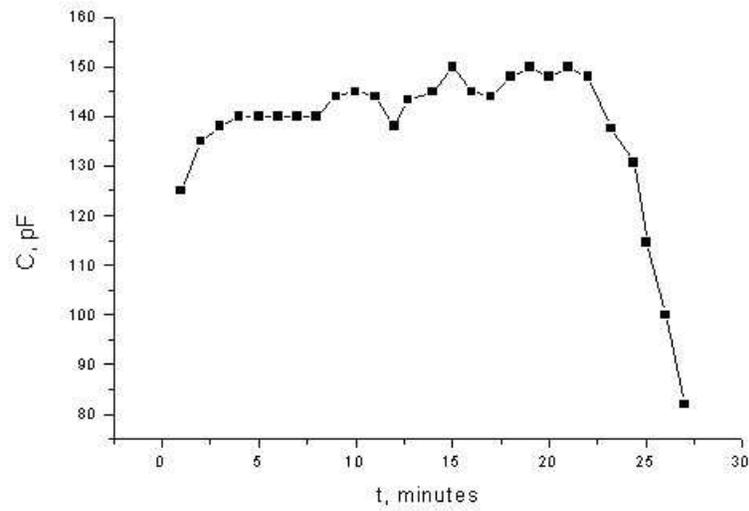}
\caption{\small{Capacity of the 50\%-aqueous solution of alcohol as a
function of time at the frequency of measuring electric field $f_{\rm meas}$ =
100 Hz without the influence of the TC but with the presence of an imitator.}} \label{Figure 4}
\end{center}
\end{figure}

\begin{figure}
\begin{center}
\includegraphics[scale=0.6]{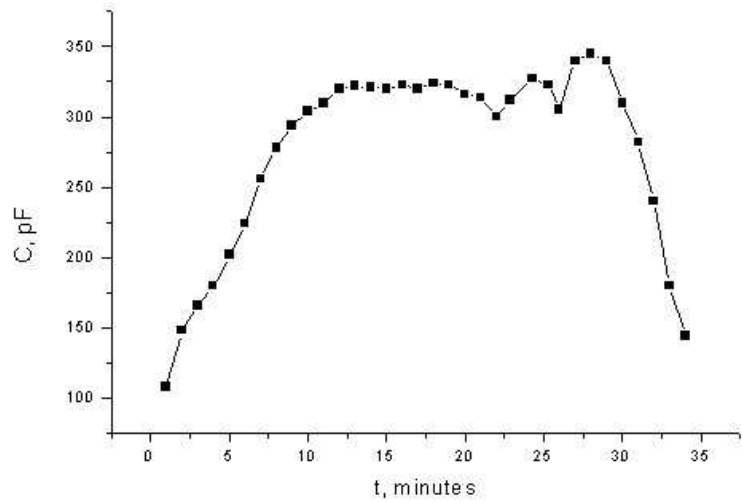}
\caption{\small{Capacity of the 50\%-aqueous solution of alcohol affected
by the TC as a function of time at the frequency of measuring electric field
$f_{\rm meas}$ = 100 Hz.}} \label{Figure 5}
\end{center}
\end{figure}

\subsubsection{Measurements of scheme ({\textit b})}

In the experiments of the kind\textit{ (b)}
we used the set-up shown in Figure 6

\begin{figure}
\begin{center}
\includegraphics[scale=0.6]{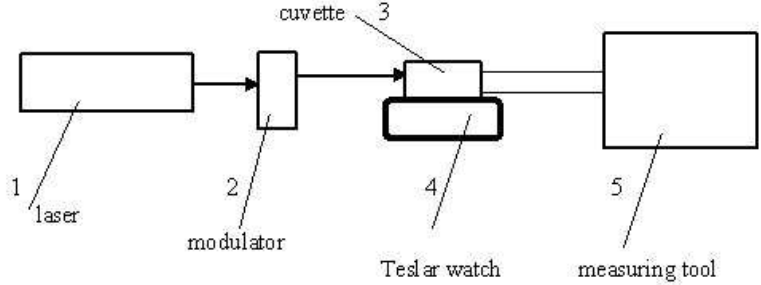}
\caption{\small{Scheme of the experiments of the kind
\textit{(b)}.}} \label{Figure 6}
\end{center}
\end{figure}

Here the measuring cell ``3'' is a cuvette with sizes indicated above. Where
the laser beam enters the cuvette, a gasket made not of teflon, but
BaF$_{2}$, which is transparent to the laser beam with the wavelength
$\lambda $= 0.63 $\mu$m.

The source of continuous radiation ``1'' was the gaseous He-Ne laser
$\Lambda \Gamma {\rm H}$-113, whose power parameters were controlled
by the pyroelectric tester of power $\Pi {\rm B} \amalg_{\!\!\! -}$-2,
worked out by our Institute of Physics. The instrument certified
in Ukraine measures the power in the range of 10$^{-7}$ to 1 W and
in the spectral range of 0.3 to 15 $\mu$m. In the present
experiments the power of laser beam was equal to $P = 8$ mW.

The flow of laser radiation was modulated by
the mechanical modulator ``2'',  which enters the makeup of
the power tester $\Pi {\rm B} \amalg_{\!\!\! -}$-2. The frequency
of modulation could be tuned between the range of 7 to 20 Hz
with accuracy of 0.1 Hz.

It should be particularly emphasized the significance of this
experimentation: it allows us to act upon the aqueous solution under
examination in the frequency range close to 7 to 9 Hz, which as presupposed
is distinctive for the non-specific radiation of the TC.

The capacity of the solution has been measured by device ``5'', the
measuring tool of impedance E7-15. The same characteristics of this
measuring tool as described in experiment ({\it a}) were exploited. Conditions at
which our experiments were conducted, the technique and equipment set-up,
were within the National Standards of Ukraine.

In the experiments, we have investigated how the capacity of aqueous
solution varies with time both under the influence of the TC, and outside
the influence of the TC. The watch ``4'' was placed as shown in Figure 6.

The distances between parts of the set-up shown in Figure 6 are the
following. The distance from the exit window of laser ``1'' to the modulator
``2'' was about 30 cm; the cuvette ``3'' was divided from the modulator
``2'' by 25 cm; the Teslar watch ``4'' was separated by 1 mm from the
cuvette ``3''; the measuring tool ``5'' was separated by 50 cm from the
cuvette ``3''.

The experimental results obtained are presented in Figures 7 and 8.

\begin{figure}
\begin{center}
\includegraphics[scale=0.65]{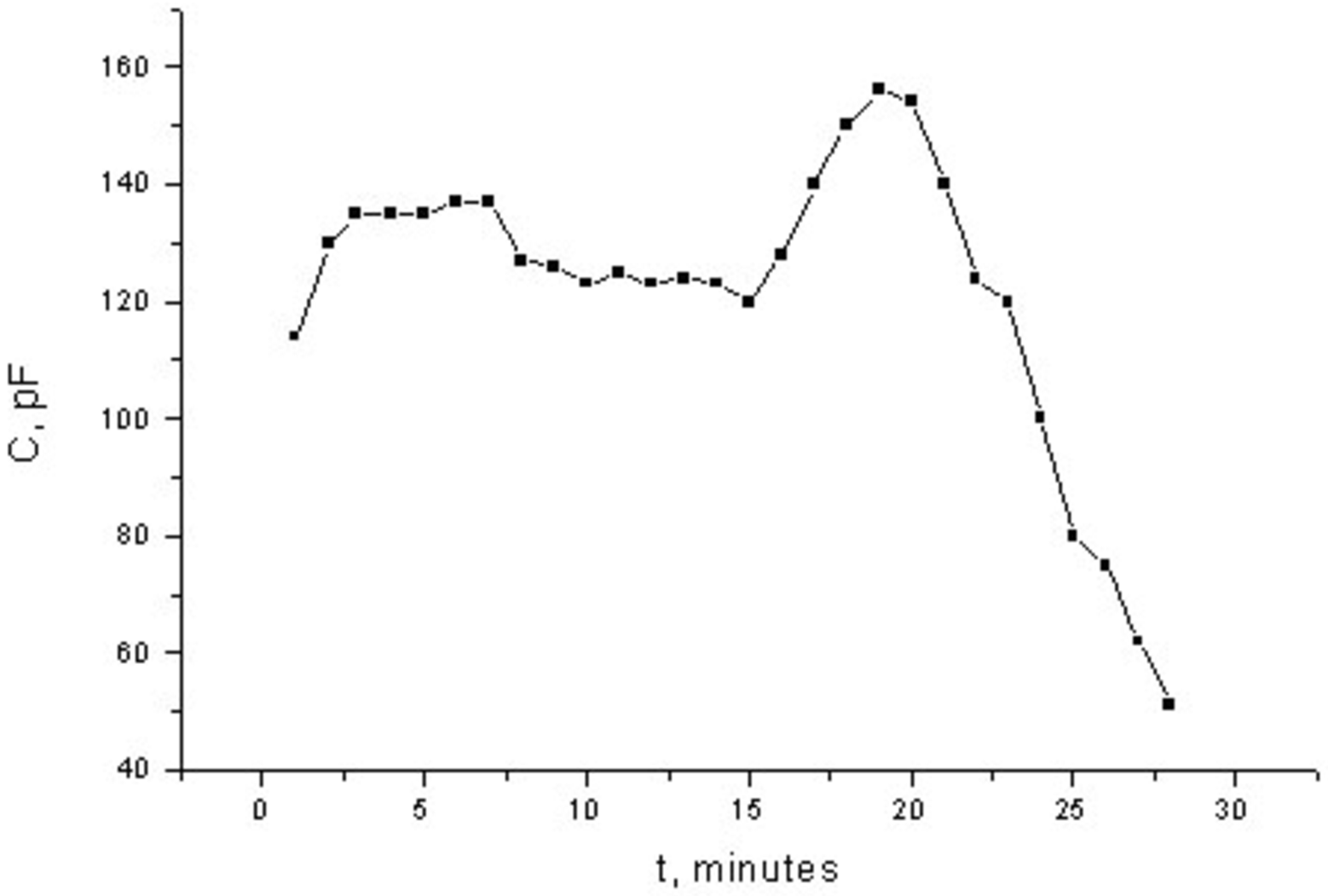}
\caption{\small{Capacity of the 50\%-aqueous solution of alcohol as
a function of time, without the TC, but in the presence
of an imitator. The cuvette is scanned by the measuring
electric field with the frequency $f_{\rm meas} = 1$ kHz
and is irradiated by the laser beam. The frequency of
mechanical modulation $f_{\rm mod}$ of the laser beam
changes as follows: $f_{\rm mod} = 7$ Hz from the
1$^{\rm st}$ to 10$^{\rm th}$ minutes;
$f_{\rm mod} = 8$ Hz from 11$^{\rm th}$ to
13$^{\rm th}$ minutes; $f_{\rm mod} = 9$ Hz from
14$^{\rm th}$ to  19$^{\rm th}$ minutes; and
$f_{\rm mod}$ = 10 Hz from 20$^{\rm th}$ to 28$^{\rm th}$
minutes.}} \label{Figure 7}
\end{center}
\begin{center}
\includegraphics[scale=0.6]{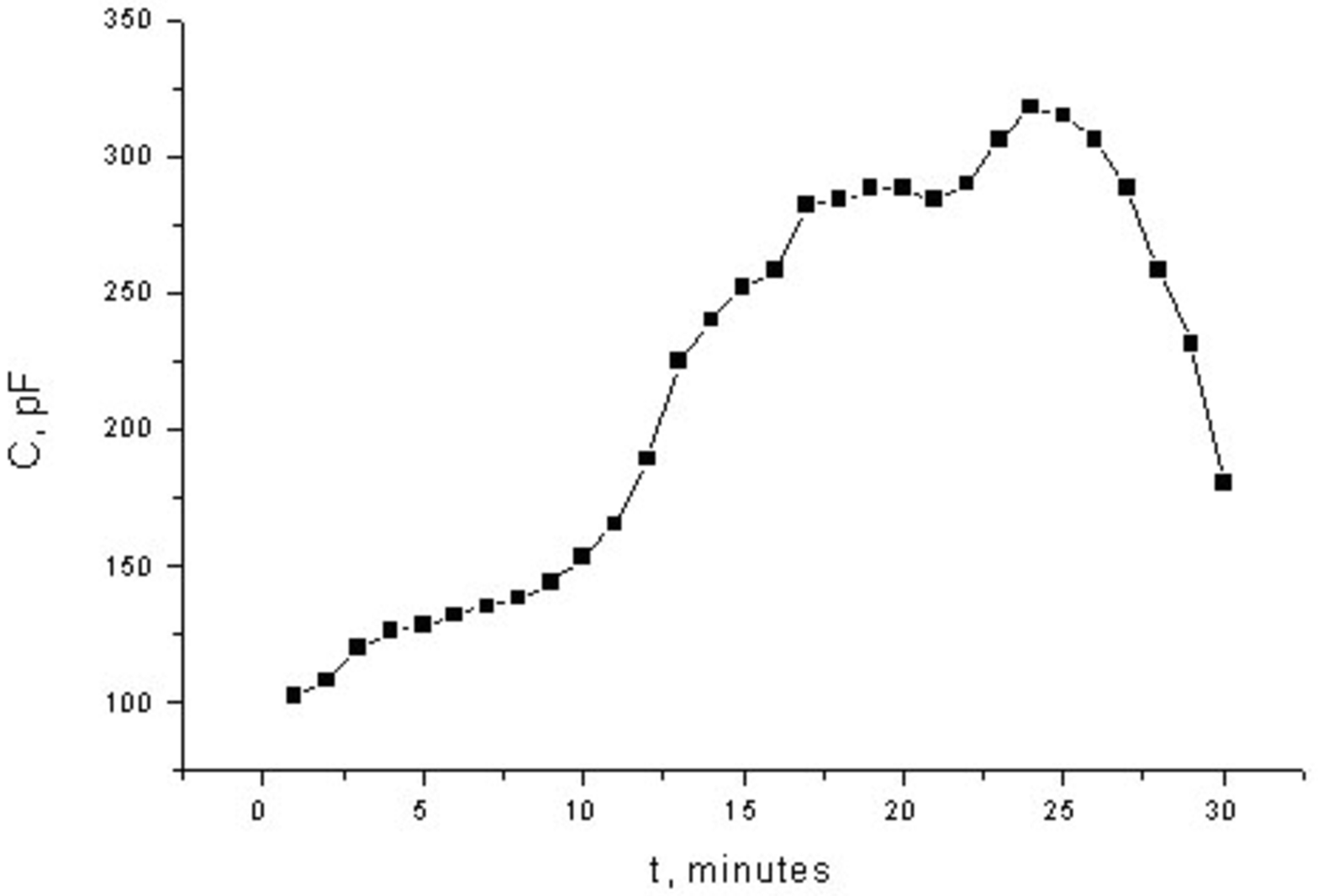}
\caption{\small{Capacity of the 50\%-aqueous solution of alcohol
affected by the TC as a function of time. The cuvette is
scanned by the measuring electric field with the frequency
$f_{\rm meas} = 1$ kHz and is irradiated by the laser beam.
The frequency of mechanical modulation
$f_{\rm mod}$ of the laser beam changes as follows:
$f_{\rm mod} = 7.6$ Hz from the 1$^{\rm st}$ to
8$^{\rm th}$ minutes; $f_{\rm mod} = 8$ Hz during the
9$^{\rm th}$ minute;  $f_{\rm mod} = 9$ Hz during the
10$^{\rm th}$ minute; $f_{\rm mod} = 10$ Hz from
11$^{\rm th}$ to 16$^{\rm th}$ minutes; and
$f_{\rm mod} = 20$ Hz from 17$^{\rm th}$ to
30$^{\rm th}$ minutes.}} \label{Figure 8}
\end{center}
\end{figure}

\subsection{Discussion of experimental results}

The results of measurements presented above show an unusual behavior of
dielectric properties of the 50\%-aqueous solution of alcohol.

The key results of the experiments can briefly be stated as follows. Under
the action of radiation of the TC the alcohol component is evaporated more
intensively from the solution; this is evident from the study of the
solution density. At the same time the water component is specified by a
``frozen'' state.

These results are associated with the behavior of the permittivity of the
solution, because the capacity \textit{C} is proportional to $\varepsilon $
and the permittivities of the solution components are $\varepsilon _{{\kern
1pt} {\rm w}} = 81$ (water) and $\varepsilon _{{\kern 1pt} {\rm a}} = 26$ (alcohol).
The real part of the permittivity $\varepsilon $ of the solution increases
remarkably with time, which is seen from Figures 3 and 5. The following
decrease of  $\varepsilon $ is provoked by the evaporation of the remaining water.
This is apparent from the experiments of the kind \textit{(a)}.

A very similar behavior shows the aqueous solution of alcohol in the case of
the experiments of the kind \textit{(b)}, Figure 8. However, here, the
change of $\varepsilon $ strongly depends on the value of the modulating
frequency $f_{\rm mod}$ of laser beam. In the range $7.6 \, {\rm Hz} \, \le
\, f_{{\kern 1pt} {\rm mod}}   \le  9 \,{\rm Hz}$ we observe a
very peculiar restraining of the evaporation process. The increase of the
value of $f_{\rm mod}$ leads to the increase of the real part of the
permittivity $\varepsilon $, which in turn means the intensification of
evaporation of the alcohol component.

The observed anomalies of the behavior of $\varepsilon $ cannot be explained
in the framework of classical electrodynamics in principle. Indeed, the
capacity of our measuring cell filled with the aqueous solution has to be
determined as the appropriate expression written for the plane capacitor,
\begin{equation}
\label{eq1}
C = \frac{{\varepsilon _{0} {\kern 2pt}
\varepsilon {\kern 1pt} {\kern 1pt} S}}{{d}},
\end{equation}

\noindent
where $\varepsilon _{0}$ is the dielectric constant,
$\varepsilon$ is the permittivity of the substance in
question, \textit{d} is the width of the capacitor and $S$ the square
of capacitor plate. One can say that at initial stages of measurements the
increase in capacity \textit{C} can be associated with the increase of
$\varepsilon $ of the aqueous solution. The rate of this increase exceeds
the rate of the decrease of the square \textit{S}. The value of \textit{S},
which is defined by the quantity of aqueous solution that fills the
capacitor, gradually diminishes due to several reasons: the natural
evaporation of the alcohol component and the irradiation by the measuring
field and the laser beam.

The relative dielectric constant of the 50\%-aqueous solution of alcohol can
be written in the form (see, e.g. Ref. 4)
\begin{equation}
\label{eq2}
\varepsilon = \frac{{m_{{\kern 1pt} a}} }{{m_{{\kern 1pt} a} + m_{{\kern
1pt} w}} }{\kern 1.5 pt}\varepsilon _{{\kern 1pt} a} \,\, + \,\,\,
\frac{{m_{{\kern 1pt} w} }}{{m_{{\kern 1pt} a} + m_{{\kern 1pt} w}} }
{\kern 1.5 pt}\varepsilon _{{\kern 1pt} w}
\end{equation}

\noindent
where $m_{{\kern 1pt} {\rm w}} $ and $m_{{\kern 1pt} {\rm a}}$
are the water and alcohol masses in the aqueous solution,
respectively; $\varepsilon _{{\kern 1pt} {\rm w}}$
and $\varepsilon _{{\kern 1pt} {\rm a}} $ are
the relative dielectric constants of water and alcohol,
respectively.

At the initial condition, the dielectric constant
$\varepsilon$ of the 50\%-solution can be presented
by the following expression
\begin{equation}
\label{eq3}
\varepsilon = 0.5{\kern 1pt} {\kern 1pt} \varepsilon _{{\kern 1pt} w} +
0.5{\kern 1pt} {\kern 1pt} \varepsilon _{{\kern 1pt} a} = 0.5{\kern 1pt}
{\kern 1pt} {\kern 1pt} \times {\kern 1pt} {\kern 1pt} {\kern 1pt} 81 +
0.5{\kern 1pt} {\kern 1pt} {\kern 1pt} \times {\kern 1pt} {\kern 1pt} {\kern
1pt} 27 = 54
\end{equation}

\noindent
where $\varepsilon _{{\kern 1pt} {\rm w}} = 81$ and
$\varepsilon _{{\kern 1pt} {\rm a}} = 27$.

Owing to the evaporation, the maximum possible value of $\varepsilon $ can
reach is the value $\varepsilon = \varepsilon _{{\kern 1pt} {\rm w}} = 81$ (when
the alcohol component is completely evaporated). Due to the evaporation, the
liquid surface \textit{S} of the capacitor decreases by a factor of two,
$S/2$. Therefore, one can anticipate the reduction of the capacity by
25 \%. However, contrary to the anticipation, we have observed a significant
increase in the capacity, up to 5.5 times (see Figure 3). Clearly such a
phenomenon must be associated only with the influence of the TC on the
aqueous solution.

\section{The Theory of the Phenomenon}

\subsection{Phenomenological consideration}

The expression for dielectric permittivity
$\varepsilon$ can be written in the following general way
\begin{equation}
\label{eq4}
\varepsilon = \frac{{d{\kern 1pt} D}}{{d{\kern 1pt} E}},
\end{equation}

\noindent
where \textit{E} and \textit{D} are the electric field strength and the
electrical induction of the solution studied, respectively. In the linear
approximation
\begin{equation}
\label{eq5}
D = \varepsilon _{0}{\kern 1pt} E + P
\end{equation}

\noindent
where \textit{P} is the polarization of molecules, which is
represented as the sum of electronic $P_{\rm e}$,
nuclear $P_{\rm n}$ and orientational $P_{\rm o}$ components \cite{5}
\begin{equation}
\label{eq6}
P = P_{\rm e} + P_{\rm n} + P_{\rm o} .
\end{equation}

With an intensive evaporation, the structure of the aqueous solution of
alcohol changes and obviously the major contribution to a change of
\textit{P} introduces the orientational polarization $P_{\rm o} $. Let us
consider the dependency of $P_{\rm o} $ on the side of the scalar field
generated by the TC representing it as a superposition of two
electromagnetic waves whose amplitudes are the same but are displaced in
phase by 180$^{\circ}$, such that the electromagnetic polarization becomes
compensated. In what way is such a resultant field able to interact with
chaotically oriented dipole moments of molecules of the solution?

It is well-known that in an applied field $\vec {E}$ the potential energy
\textit{U} of a molecule with a dipole moment $\vec {p}$ is equal to
\begin{equation}
\label{eq7}
U = - \vec {p} \cdot \vec {E} = p{\kern 1.5pt}E{\kern 1pt}\cos\vartheta
\end{equation}

\noindent
where $\vartheta $ is the angle between vectors $\vec {p}$ and $\vec {E}$.
If one changes the direction of the electric field $\vec {E}$ to the
opposite one, $ - \vec {E}$, then $\cos\vartheta$ must also change the sign.
Therefore, two waves having opposing polarization, which are phase
shifted by 180$^{\circ}$, will be characterized by the doubled
potential energy (\ref{eq7}), i.e. the energy of a dipole in these
two waves will be specified by the value $2{\kern 1pt} U$.
Then following Ref.~{5}, we can write the resulting orientation
polarization of the solution studied
\begin{equation}
\label{eq8}
P_{\rm o} = n p{\kern 1pt} {\kern 1pt} L\left( {2p{\kern 1pt} E/k_{\rm B}
T} \right)
\end{equation}

\noindent
where $k_{\rm B} $ is the Boltzmann constant, \textit{T} the temperature,
\textit{n} the concentration of molecules and $L\left( {x} \right)$ is the
Langeven function determined as
\begin{equation}
\label{eq9}
L\left( {2p{\kern 1pt} E/k_{\rm B} T} \right) = \coth\left( {2p{\kern 1pt}
E/k_{\rm B} T} \right){\kern 1pt}  -
\,\,k_{\rm B} T/2p{\kern 1pt} E
\end{equation}

At the room temperature, which corresponds to the experimental conditions,
$2p{\kern 1pt} E \ll  k_{\rm B} T$ and hence $L\left( {2p{\kern 1pt} E/k_{\rm B} T}
\right) \approx 2p{\kern 1pt} E/\left( {3k_{\rm B} T} \right)$. That is why the
polarization becomes
\begin{equation}
\label{eq10}
P_{\rm o} = 2n{\kern 1pt} p{\kern 1pt} ^{2}E/\left( {3{\kern 1pt} {\kern 1pt}
k_{\rm B} T} \right).
\end{equation}

In expression (\ref{eq10}) the value of the field $E$
should represent the time-averaged component of the scalar field.
Averaged effective amplitude $E$ of two superposed electric fields
can easily be derived,
\begin{equation}
\label{eq11}
E = \bar {E} + \bar {E} = 2 \times \frac{1}{T/2}\int\limits_{0}^{T/2}
{E_{0}}  \sin (\omega {\kern 1pt} t)  {\kern
1.5pt}  d{\kern 1pt} t = \frac{4}{\pi}{\kern 1pt} E_{0}
\end{equation}

\noindent
where $\omega = 2{\kern 1pt} {\kern 1pt} \pi /T$ and $E_{0} $ is the
amplitude of each of the two electromagnetic waves.

Thus we may suggest that the frequency of the scalar field is twice as
larger in comparison with the frequency of each electromagnetic components,
whose superposition forms the scalar wave in question. Therefore, expression
(\ref{eq10}) becomes
\begin{equation}
\label{eq12}
P_{\rm o} = 8{\kern 1pt} n{\kern 1pt} p^{\kern 1pt 2}E_{0} /\left( {3\pi
{\kern 1pt} k_{\rm B} T} \right).
\end{equation}

A similar reasoning has to be true for the measuring field. That is, the
contribution to $\varepsilon $ on the side of the scalar field studied,
which is determined as $\Delta {\kern 1pt} {\kern 1pt} \varepsilon = \Delta
{\kern 1pt} P_{\rm o} /\Delta {\kern 1pt} E$, should be 4 times a conventional
measuring field.

The large value of polarization (\ref{eq12}) means a substantial growth of the
dielectric permittivity. In other words, the scalar field ``freezes''
dipoles of water molecules, which results in the increase of the
polarization $P_{\rm o} $ of the aqueous solution of alcohol.

\subsection{Submicroscopic foundations}

In the TC two flows of electromagnetic field, which spread in the same
direction, are canceled and this creates a scalar low frequency wave that
continues to transfer the energy stored in the electromagnetic field. That
was the hypothesis of the authors of the invention.

One can ask whether this is possible. Although this would be accounted for
in phenomenological terms, as has been described in the previous subsection,
the answer of the conventional foundations of physics, which is based on
orthodox quantum theory, is rather negative, because these foundations are
not fundamental enough. At the same time, the submicroscopic concept of the
foundations of physics [6-9] allows us
to account for the cancellation of two electromagnetic waves that
spread along the same line and whose electric
(and magnetic) polarization is shifted on the phase 180$^{\circ}$.

In Ref.~{10} and Ref.~{11} a detailed theory of
the photon, which is an elementary carrier of electromagnetic
waves, and the electric charge as such have been
developed starting from first submicroscopic principles
that consider the real physical space as a tessellation lattice
(called the tessellattice) of primary elements, cell,
balls, or superparticles. It has been argued
that the electric and magnetic polarizations of a photon
are associated with the surface profile of the photon,
i.e. the electric and magnetic polarization are played on
the surface of the primary cell of the
tessellattice and this polarization is transferred
from cell to cell by relay mechanism.

One more excitation of the tessellattice is the inerton:
this is a local excitation of the tessellattice,
located in a cell and which moves by relay
mechanism, as well. The inerton is associated with
the reduction of the volume of a cell.
This is a mass excitation.

A photon is also characterized by the volume reduction,
because this is the initial condition for an excitation to exist.
Initially a fractal volumetric deformation emerges in
a degenerate cell of the tessellattice and this is
the mass excitation (see Ref.~{7} and Ref.~{8}).
The state of the surface of this excitation determines whether
it is: 1) a pure mass excitation (the
inerton), which does not have any surface polarization, or
2) a mass excitation that additionally is characterized by
a special relief of the cell surface (the photon).

Thus if two photons, which are opposite in phase
and are spreading along the same line touch each other,
the polarizations pertaining to the photon
surfaces should neutralize one another.
This brings about neutral massive
excitations of the space, i.e. inertons.

By this means it is quite possible the emergence of
inertons owing to the cancellation of photons.
These inertons will continue to move along the path
of initial photons and will transfer the same energy
that had been carried by photons.

\subsection{Collective interaction of water molecules}

In the case of the TC, setting for the frequency of the TC's
inertons $\nu \approx 8$ Hz, we can write the relations
\begin{equation}
h\nu = m{\kern 1pt} {\kern 1pt} \tilde {c}^{\kern 1pt 2} = 5.3
\times 10^{ - 38} \ {\rm J}
\label{13}
\end{equation}

\noindent
where \textit{m} and $\tilde {c}$ are the mass and
the velocity of the inerton.

As has been shown in Ref.~{12}, the vibratory potential
of water molecules, which is associated with their elastic
properties, is caused by the overlapping of inerton
clouds of vibrating entities. This means that
vibrating entities, which interact through the inerton
interaction (a subsystem of the matter waves),
are able to fall into the interaction with
an applied inerton field. Thus, in the aqueous solution,
molecules of H$_{2}$O and C$_{2}$H$_{5}$OH engage
with the external inerton field radiated by the TC
(i.e. a flow of mass) and become the receptors of the
inerton radiation.

The most interesting is the water molecule H$_{2}$O, because it can be
treated as both the ``mass dipole'' and the electric dipole. In fact, the
water molecule is asymmetric: one edge is heavy (oxygen) and the other edge
is light (two hydrogen). Hence the heaviest edge should turn to the source
of the inerton radiation (i.e. the TC) and the light edge should be oriented
in the opposite direction. The electric dipole exactly superimposes on this
``mass dipole'' (oxygen has the negative charge and a pair of hydrogen has
the positive charge).

It is important to note, the measuring electric field can introduce
microscopic perturbations in the samples studied; in our case, the intensity
of the measuring field was not small enough to avoid this; its intensity was
in the range 10 to 100 mW/cm$^{2}$. Therefore among other flows in the
cuvette one can distinguish a micro conventional flow that is characterized
by the same value of frequency, namely, $\nu \approx 8$ Hz. This means in
light of submicroscopic mechanics developed by
Krasnoholovets [13-16] that in this flow,
water molecules should be specified by the following kinetic
parameters: the de Broglie wavelength of the molecule
$\lambda \sim 10^{- 5}$ m, the molecule velocity
$\upsilon \sim 10^{{\kern 1pt} - 3}$ m/s
and the frequency of spatial oscillations of
the particle $\nu = 2{\kern 1pt} \upsilon /\lambda$ \break
$ \approx 8$ Hz.

If the inerton field radiated by the TC orders water molecules, we can then
suggest that ordered water molecules begin to interact stronger. In other
words, this should result in additional correlation between dipoles of water
molecules. Figure 9 discloses this mechanism graphically.

Water molecules with these parameters begin to interact resonantly with the
inerton field of the TC. Or more exactly, the inerton field stimulates all
the water molecules in the cuvette to synchronic motion with the
aforementioned parameters.

Let us put for the estimate the smallest value of the intensity of
the inerton field that the TC can radiate, $I = 10^{- 6}$ W/cm$^{2}$.
Then in time $t = 100$ s each water molecule receives
a dose of inerton energy equal to
\begin{equation}
\label{eq14}
 E = I{\kern 1pt}t{\kern 1pt}\sigma = 10^{{\kern 1pt} - 6} \
 {\rm W/cm^2} \, \times \,\,100\;{\rm s} \, \times 10^{ - 16}
 \;{\rm cm^2} = 10^{{\kern 1pt} - 20} \ {\rm J}
\end{equation}

\noindent
where $\sigma = 10^{ - 16}$ cm$^{2}$ is the true cross-sectional area of a
water molecule (recall the thermal energy at the room temperature is $k_{\rm B}
T = 4.25 \times  10^{ - 21}$ J). The energy (14) is used for the further
generation of the synchronic motion of water molecules and partly the energy dissipates.

\begin{figure}
\begin{center}
\includegraphics[scale=0.9]{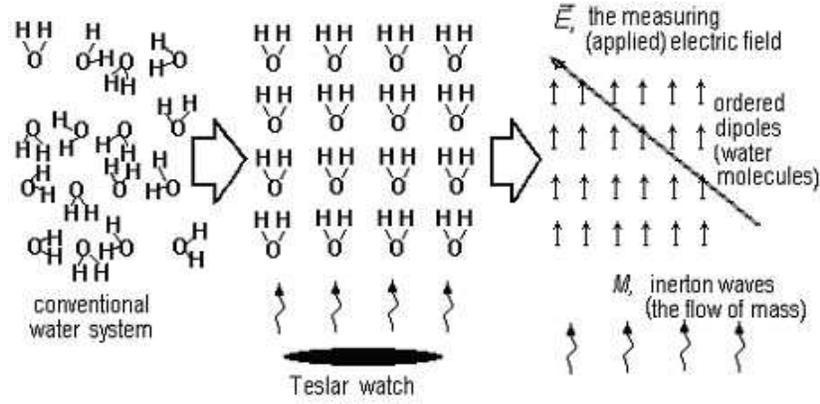}
\caption{\small{Behavior of the dynamic water system affected by the TC.}} \label{Figure 9}
\end{center}
\end{figure}

Since the inerton cloud surrounding each molecule is exemplified by the
radius [13-15]
\begin{equation}
\label{eq15}
\Lambda = \lambda {\kern 1pt} {\kern 1pt} \tilde {c}/\upsilon ,
\end{equation}

\noindent
inertons of each molecule completely cover all the other molecules in the
cuvette (this follows from the meanings of the parameters above). On the
other hand, the inerton cloud of a water molecule represents exactly
the range of space topped by the wave $\psi$-function that is employed in the
orthodox quantum mechanical formalism. The overlapping of inerton clouds of
water molecules allows the transference of absorbed inertons irradiated by
the TC to other molecules in the aqueous solution. During time \textit{t} =
100 s a water molecule can absorb $N = E/(h{\kern 0.5pt}\nu) \sim 10^{12}$
inertons. Then the mass of this molecule should obey the kinetic equation
\begin{equation}
\label{eq16}
\dot {m} = - \alpha {\kern 1pt} m + \beta M
\end{equation}

\noindent
where the term $\alpha {\kern 1pt} m$ describes the dissociation and the term
$\beta {\kern 1pt} {\kern 1pt} M$ depicts the growth of the molecule mass
(due to the absorption of the mass $M$ from the TC). The solution to
equation (\ref{eq16}) has the form
\begin{equation}
\label{eq17}
m = {\kern 1pt} m_{0} {\kern 1pt} e^{ - \alpha {\kern 1pt} t} +
\beta M
\end{equation}

It is obvious that the behavior of absorbed energy $E$ (\ref{eq14}) should also
follow the rule for the mass $m$ (\ref{eq17}) and, as it follows from our
experiments, the relaxation time $1/\alpha $ can reach several tens of
minutes.

The energy (14), which a molecule absorbs, expands to other molecules
through the cloud of inertons. At the moment \textit{t} = 100 s the energy
(14) allocated to the ensemble of molecules already exceeds the thermal
energy, $k_{\rm B} T$. Note this is the potential energy, i.e. the molecule
re-allocating absorbed inertons induces a deformation of space, which
then imbibes other water molecules. This signifies the appearance of an
order parameter in the aqueous solution: Water molecules start to orientate
themselves along the lines of the inerton field of the TC.

In the language of conventional physics this means that
in the time $t \sim 100 \, \, {\rm s}$
the TC's field forms a potential well \textit{W}
for each water molecule in the cuvette and
the depth of the well is no less than the value of the
thermal energy, i.e. $|W|\,\sim  \,|E|\, \ge \, k_{\rm B} T$.

The filament ordering of water molecules caused by the external inerton
field should induce their additional dipole-dipole interaction in the same
filament. Then in expression (\ref{eq8}) the dipole moment \textit{p} of a water
molecule should be replaced by a more complicated expression that in the
approximation of nearest neighbors can be presented in the form
\begin{equation}
\label{eq18}
p \to p \times \left( {1 + \chi ( I )\,M /k_{\rm B} T} \right)
\end{equation}

\noindent
where the matrix element of the energy of interaction of
a pair of dipoles \cite{17}
\begin{equation}
\label{eq19}
M = \frac{{\sqrt {{\raise0.5ex\hbox{$\scriptstyle
{2}$}\kern-0.1em/\kern-0.15em\lower0.25ex\hbox{$\scriptstyle {3}$}}}
}}{{4{\kern 1pt} {\kern 1pt} \pi {\kern 1pt} \varepsilon _0
}}{\kern 1pt} {\kern 1pt} {\kern 1pt} \frac{{|\vec {p}_{{\kern 1pt}\rm a}
||\vec {p}_{\kern 1pt \rm b} |}}{{r_{{\kern 1pt} \rm a{\kern 1pt} \rm b}^
{{\kern 1pt} 3}} };
\end{equation}

\noindent
here $r_{{\kern 1pt}{\rm  a{\kern 0.4pt} b}}$ is the distance
between the nearest dipoles a and b in the ``filament''.
The dimensionless function $\chi \left( {I} \right)$
in expression (\ref{eq19}) should be treated as
a coupling parameter that makes an allowance for
the influence of the external inerton field on
the interaction of dipoles.

Substituting \textit{p} from expression (\ref{eq18})
(with regard to (\ref{eq19})) into expression for
the permittivity we finally obtain
\begin{equation}
\label{eq20}
\varepsilon = \frac{{n {\kern 1pt} p^{\kern 1pt 2}}}
{{3{\kern 1pt} k_{\rm B} T}} \times \left( {1 +
\sqrt {{\raise0.5ex\hbox{$\scriptstyle
{2}$}{\kern 1pt}\kern-0.1em/\kern-0.15em\lower0.25ex
\hbox{$\scriptstyle {3}$}}} {\kern 3pt}
\chi (I)\,\frac{{p^{\kern 1pt 2}}}
{{4{\kern 1pt} {\kern 1pt} \pi {\kern 1pt}
\varepsilon _{0} {\kern 1pt} g^{3}}}} \right)^{2}
\end{equation}

\noindent
where the indices are omitted and the designation for the lattice constant
\textit{g} (the distance between the nearest molecules) is introduced,
\textit{p} is the dipole moment of water molecule, \textit{n} is the
concentration of water molecules.

Let us assign now numerical values to the parameters
in expression (\ref{eq20}): \break
$p = 6.2 \times 10^{ - 30}$ C$\cdot$m, $g = 0.281$ nm,
$k_{\rm B} T = 4.25 \times 10^{- 21}$ J and
$\varepsilon _{0} = 8.85 \times 10^{ - 12}$ F/m.
If we put for the coupling parameter
$\chi \left( {I} \right) = 0.45,$ we can easy
calculate the expression in the parentheses to
the power 2; the outcome of the calculation is
equal to 5.5, which exactly corresponds to
the experimental result discussed above
(once again, see Figure 3).

\section{Conclusion}

Our experimental results show that in a water system exposed
to the Teslar$^{\scriptsize{\circledR}}$ technology, a substantial increase
of the permittivity occurs. The radiation of the
Teslar$^{\scriptsize{\circledR}}$ technology ``freezes''
dipole water molecules, which leads to the induction of
an additional value of the dipole moment in a water molecule.

We also have proposed a theory of this interesting phenomenon.
The theory is very new and is based on submicroscopic principles
of the constitution of nature. The submicroscopic concept
is the most fundamental one and can readily be
introduced as the basis for the orthodox quantum mechanical
formalism  [6-16].
This concept could already explain some
other unusual physical effects
(see in Ref.~{7} and Ref.~{14}).
This allows us to state that the submicroscopic concept
lends credibility to the theoretical analysis of
experimental data obtained in section 3 of
the present report.


{\small
\begin{thebibliography}{0}


\bibitem{1} G. Rein, {\it JUS Psychotronics Association} {\bf 1}, 15 (1989).

\bibitem{2} G. Rein, in {\it The Proc. 7th Int. Association of
Psychotronics Research}, Georgia, Dec. 1988.

\bibitem{3} A. S. Morley, {\it Health Consciousness} {\bf 11}, June,
41 (1991).

\bibitem{4} Ya. Yu. Akhadov, {\it Dielectric properties of
binary solutions} (Nauka, Moscow, 1977), in Russian.

\bibitem{5} C. Kittel, {\it Introduction to solid state physics}
7th edn. (John Wiley, New York, 1996).

\bibitem{6} M. Bounias and V. Krasnoholovets,
{\it Kybernetes: The Int. J. Systems and Cybernetics}
{\bf 32}, 945 (2003) (also physics/0211096).

\bibitem{7} M. Bounias and V. Krasnoholovets, {\it ibid.}
{\bf 32}, 976 (2003) (also physics/0212004).

\bibitem{8} M. Bounias and V. Krasnoholovets, {\it ibid.} {\bf 32},
1005 (2003) (also physics/0301049).

\bibitem{9} M. Bounias and V. Krasnoholovets,  {\it Int.
J. Anticipatory Computing Systems} {\bf 16}, 3 (2004) (also
physics/0309102).

\bibitem{10} V. Krasnoholovets, {\it Ann.
Fond. L. de Broglie} \textbf{27}, 93 (2002) (also
quant-ph/0202170).

\bibitem{11} V. Krasnoholovets,
{\it Hadronic J.} {\bf 18}, 425 (2003)
(also physics/0501132).

\bibitem{12} V. Krasnoholovets, {\it Central Eur. J.
Phys.} {\bf 2}, 698 (2004).

\bibitem{13} V. Krasnoholovets and D. Ivanovsky,
{\it Phys. Essays} {\bf 6}, 554 (1993) (also
quant-ph/9910023).

\bibitem{14} V. Krasnoholovets, {\it Phys. Essays}
{\bf 10}, 407 (1997) (also quant-ph/9903077).

\bibitem{15} V. Krasnoholovets, {\it Ind. J. Theor. Phys.}
{\bf 48}, 97 (2000) (also quant-ph/0103110).

\bibitem{16} V. Krasnoholovets,
{\it Int. J. Computing Anticipatory Systems}
{\bf 11}, 164 (2002) (also quant-ph/0109012).

\bibitem{17} V. M. Agranovich and M. D. Galanin,
{\it The transfer of energy of electronic excitation
in condensed media} (Nauka, Moscow, 1978), p. 27; in
Russian.

\end{thebibliography} }
\end{document}